\documentclass[journal,twoside,web]{ieeecolor}
\usepackage{generic}
\usepackage{cite}
\usepackage{amsmath,amssymb,amsfonts}
\usepackage{algorithmic}
\usepackage{graphicx}
\usepackage{hyperref}
\usepackage{textcomp}
\usepackage{tabularx}
\usepackage{multirow}

\def\BibTeX{{\rm B\kern-.05em{\sc i\kern-.025em b}\kern-.08em
    T\kern-.1667em\lower.7ex\hbox{E}\kern-.125emX}}
\begin{document}

\title{Diff-DTI: Fast Diffusion Tensor Imaging Using A Feature-Enhanced Joint Diffusion Model}
\author{Lang Zhang, Jinling He, Dong Liang, \IEEEmembership{Senior Member, IEEE}, Hairong Zheng, \IEEEmembership{Senior Member, IEEE}, Yanjie Zhu, \IEEEmembership{Senior Member, IEEE}
\thanks{This work was supported in part by the National Natural Science Foundation of China under Grant 63232119, 12226008, 62125111, U1805261, 62106252, 62206273, and 62201561, in part by the National Key R\&D Program of China under Grant 2020YFA0712200 and Grant 2021YFF0501402, in part by the Guangdong Basic and Applied Basic Research Foundation under Grant 2023B1212060052, and in part by the Shenzhen Science and Technology Program under Grant RCYX2021060910444089 and JCYJ20220818101205012. (Corresponding author: yj.zhu@siat.ac.cn (Yanjie Zhu).)}
\thanks{Lang Zhang, Jinling He, Dong Liang, Hairong Zheng and Yanjie Zhu are with Lauterbur Research Center for Biomedical Imaging, Shenzhen Institute of Advanced Technology, Chinese Academy of Sciences, Shenzhen, China.}
\thanks{LangZhang is also with the Chongqing university of technology, Chongqing, China.}
\thanks{Jinling He is also with the Lanzhou Jiaotong University, Lanzhou, China.}
\thanks{Yanjie Zhu is also with the National Center for Applied Mathematics
Shenzhen (NCAMS), Shenzhen 518000, China.}}
\maketitle

\begin{abstract}
Magnetic resonance diffusion tensor imaging (DTI) is a critical tool for neural disease diagnosis. However, long scan time greatly hinders the widespread clinical use of DTI. To accelerate image acquisition, a feature-enhanced joint diffusion model (Diff-DTI) is proposed to obtain accurate DTI parameter maps from a limited number of diffusion-weighted images (DWIs). Diff-DTI introduces a joint diffusion model that directly learns the joint probability distribution of DWIs with DTI parametric maps for conditional generation. Additionally, a feature enhancement fusion mechanism (FEFM) is designed and incorporated into the generative process of Diff-DTI to preserve fine structures in the generated DTI maps. A comprehensive evaluation of the performance of Diff-DTI was conducted on the Human Connectome Project dataset. The results demonstrate that Diff-DTI outperforms existing state-of-the-art fast DTI imaging methods in terms of visual quality and quantitative metrics. Furthermore, Diff-DTI has shown the ability to produce high-fidelity DTI maps with only three DWIs, thus overcoming the requirement of a minimum of six DWIs for DTI.
\end{abstract}


\section{Introduction}
\label{sec:introduction}
\IEEEPARstart{M}{agnetic} resonance diffusion tensor imaging (DTI) is an advanced magnetic resonance imaging technique that can reflect microstructures and pathways of neural fibers within the brain by measuring the diffusion movement of water molecules in tissues\cite{bammer2003basic}\cite{basser1994estimation}\cite{le2001diffusion}. The diffusion process is typically characterized by a diffusion tensor \(D\), from which multiple DTI maps can be derived, such as Fractional Anisotropy (FA), Mean Diffusivity (MD), and Color FA\cite{alexander2007diffusion}\cite{jiang2006dtistudio}. These DTI maps hold significant clinical value in diagnosing and monitoring brain diseases like Alzheimer's and traumatic brain injury, as well as in studying pathological conditions of other tissues\cite{wilmskoetter2022language,chen2023abnormal,harrison2020imaging,ferreira2014vivo}.

However, long scan time greatly hinders the widespread clinical use of DTI\cite{jones2004effect,jones2013white}. Theoretically, DTI requires at least seven images, comprising one baseline image (b0 image) and six diffusion-weighted images (DWIs) with varied directions, to calculate the diffusion tensor \(D\)\cite{landman2007effects}. Since accurate estimation of \(D\) is crucial for generating high-quality DTI maps, it is typical to execute over 30 DWI acquisitions across different diffusion directions to improve the estimation accuracy\cite{jones2004effect,fick2016mapl}. Meanwhile, multiple repetitions are usually employed due to the inherently low signal-to-noise ratio of DWIs, which further prolongs the scan time\cite{jones2014gaussian,basu2006rician}. The long scan time of DTI causes discomfort of patients and increases the risk of motion artifacts, particularly for specific populations, such as infants or critically ill patients\cite{aksoy2011real,jeon2018peripheral}. 

To address the above issue, fast-imaging approaches have been explored to accelerate the acquisition of DTI. These approaches can be broadly categorized into two strategies. The first involves accelerating data acquisition through undersampling in k-space, followed by employing image reconstruction techniques to recover DWIs or DTI maps from the undersampled data\cite{menzel2011accelerated,zhu2017direct, huang2019accelerating, teh2020improved, varela2023single}. Compressed sensing (CS)-based reconstructions are typically employed for this purpose, primary leveraging priors derived from the correlation between DWIs. Joint sparsity and low-rank models are two typically used priors in fast DTI\cite{waters2011sparcs}. The former explores both the sparsity in image domain and the similarity among DWIs, jointly using \(\ell_1\) and \(\ell_2\) norm to enhance the reconstruction accuracy. The low-rank model leverages the similarity between DWIs to construct the low-rank constraint. However, the construction of hand-crafted priors is not-trivial. For example, Zhu et al.\cite{zhu2017direct} explored a model-based approach with spatial and parametric constraints (MB-SPC), enhancing reconstruction accuracy through a joint sparsity prior and a diffusion tensor smoothness prior. In low-rank model research, Huang et al. \cite{huang2019accelerating} combined local low-rank priors with 3D spatial smoothing constraints to develop a novel sparse DWI reconstruction model. Current studies are integrating these two priors, utilizing the sparsity in the wavelet domain and low-rank structure priors for 7T single-shot spiral DWI imaging\cite{teh2020improved}. The second strategy reduces the number of DWIs acquired, and then generates either high-quality DWIs or DTI maps from the acquisitions using deep learning (DL) neural networks\cite{jensen2005diffusional, golkov2016q, ronneberger2015u, tian2020deepdti, li2021superdti, karimi2022diffusion, wang2024ultrafast, li2023diffusion}. For example, DeepDTI generates six high SNR DWIs from the acquired ones along with a structural T1-weighted image though a 3D convolutional neural network (CNN), thereby improving the quality of DTI maps estimated\cite{tian2020deepdti}. Moreover, DTI model can be incorporated into the network as a physical constraint to improve the network performance. Liu et.al. employed this approach to generate the unacquired DWIs from the acquired six DWIs, supplementing the loss information resulting from undersampling in diffusion directions\cite{liu2023accelerated}. The above methods still need estimate the DTI maps. An alternatively way is directly mapping DTI maps from few DWIs using a network, bypassing the additional tensor estimation step. SuperDTI and Transformer-DTI have been proposed for this purpose, utilizing CNN and Transformer, respectively\cite{li2021superdti,karimi2022diffusion}. This approach, while delivering good imaging results, fails to resolve crossing fibers in the DTI maps. Therefore, MSIS-DTI improves the problem by using T1-weighted images and multiple layers of information sharing from neighboring slices. Deep learning methods show great potential in accelerating DTI imaging\cite{wang2024ultrafast}. However, most studies did not use less than six DWI directions, and the simple structure of the designed network made it difficult to accurately capture the data distribution of the DTI maps while ignoring the correlation between different DWI directions, resulting in a loss of imaging details. Additionally, the methods of this strategy face interpretability challenges in the field of medical physics due to their ``black box'' nature\cite{huff2021interpretation}.

Recent studies have demonstrated that diffusion models have disrupted the long-standing dominance of methods such as GANs and CNNs in the image generation domain\cite{dhariwal2021diffusion,cao2024survey,yang2023diffusion}. Diffusion models are generative models that consist of a forward and a reverse process\cite{ho2020denoising,song2020score}. In the forward process, the model progressively adds noise to the clear data until it is fully transformed into a Gaussian noise distribution; in the reverse process, the model denoises by utilizing the data distribution learned during training, gradually generating a sample belong to the original data distribution from Gaussian noise. Compared with the one-step generation of CNN and GAN, the multi-step gradual generative process and robust mathematical foundation of diffusion models make them more convincing\cite{dhariwal2021diffusion}. At present, diffusion model has been widely used for task such as image generation and image-to-image translation\cite{chung2022score,song2021solving,wolleb2022diffusion,yoon2023sadm}. For example, Wolleb et al.\cite{wolleb2022diffusion} utilized a Denoising Diffusion Probabilistic Model (DDPM) combined with classifier gradients to facilitate image-to-image translation between healthy and diseased subjects across four MRI sequences. Yoon et al.\cite{yoon2023sadm} proposed a Sequence-Aware Diffusion Model (SADM) for generating longitudinal brain medical images. The model uses a sequence of ordered and timestamped images as inputs and incorporates a classifier-free sequence-aware transformer as the conditional module to achieve precise conditional generation. These applications demonstrate the power capability of diffusion model in generating complex image structures. By predicting the data distribution at various noise levels, the model effectively learns the intrinsic structure and interrelationships of the data, thus achieving complex image generation. However, for precise conditional image generation, these studies typically require additional classifier training or adopt classifier-free implicit gradients for guidance, both approaches increasing the complexity and training difficulty of the model\cite{wang2024two}. Additionally, during the reverse denoising process of diffusion models, some high-frequency information is lost\cite{cao2024high}, which leads to limited performance in images rich in detailed texture information, such as FA. It's worth noting that the application of diffusion models to fast DTI imaging has not yet been explored.

In this study, we propose a novel diffusion model for fast DTI, named Diff-DTI, which generates high-fidelity DTI maps with the guidance of one b0 and few DWIs image. Diff-DTI learns the distributional structure of multi-channel data consisting of DWIs and DTI maps by means of a joint diffusion model and uses it as a guide for generating DTI maps from Gaussian noisy images using a series of denoising steps, avoiding the need for classifiers or classifier-free conditional guidance. We introduced a Feature Enhancement Fusion Mechanism (FEFM) within Diff-DTI to enhance information fusion at the feature level, thereby improving the restoration of details in DTI maps. The FEFM designs independent Transformer-based encoder for conditional DWIs, and performs information fusion via the Diffusion Feature Fusion Block (DFB) on the multi-scale feature maps fused with global information, which effectively integrates both high-level and low-level features, and improves the capture of high-frequency detailed features by Diff-DTI. Diff-DTI was trained on the HCP dataset and compared its performance to other DTI imaging methods. The results show that Diff-DTI can generate high-fidelity DTI maps using just one b0 and three DWIs, outperforming other techniques.

\section{Methods}
The main framework of the Diff-DTI method is based on a conditional diffusion model (illustrated in Fig. 1(b)), which learns the joint diffusion of \(n\)-directional DWIs and the DTI map, \(n\le 6\). The input of the model is the pair of \((x,y)\), where \(x\) represents the DTI map and \(y\) represents the \(n\)-directional DWIs (illustrated in Fig. 1(a)). In the forward process, Gaussian noise is gradually added to \(x\) until it becomes pure noise. The reverse process generates the DTI map from noise guided by DWIs. The conditional generation is achieved by learning the score function of the joint distribution of \(p(x,y)\) using a novel neural network with FEFM shown in Fig. 2. Details of the method is introduced below.

%% fig1
\begin{figure*}[!htp]
\centerline{\includegraphics[width=0.9\textwidth]{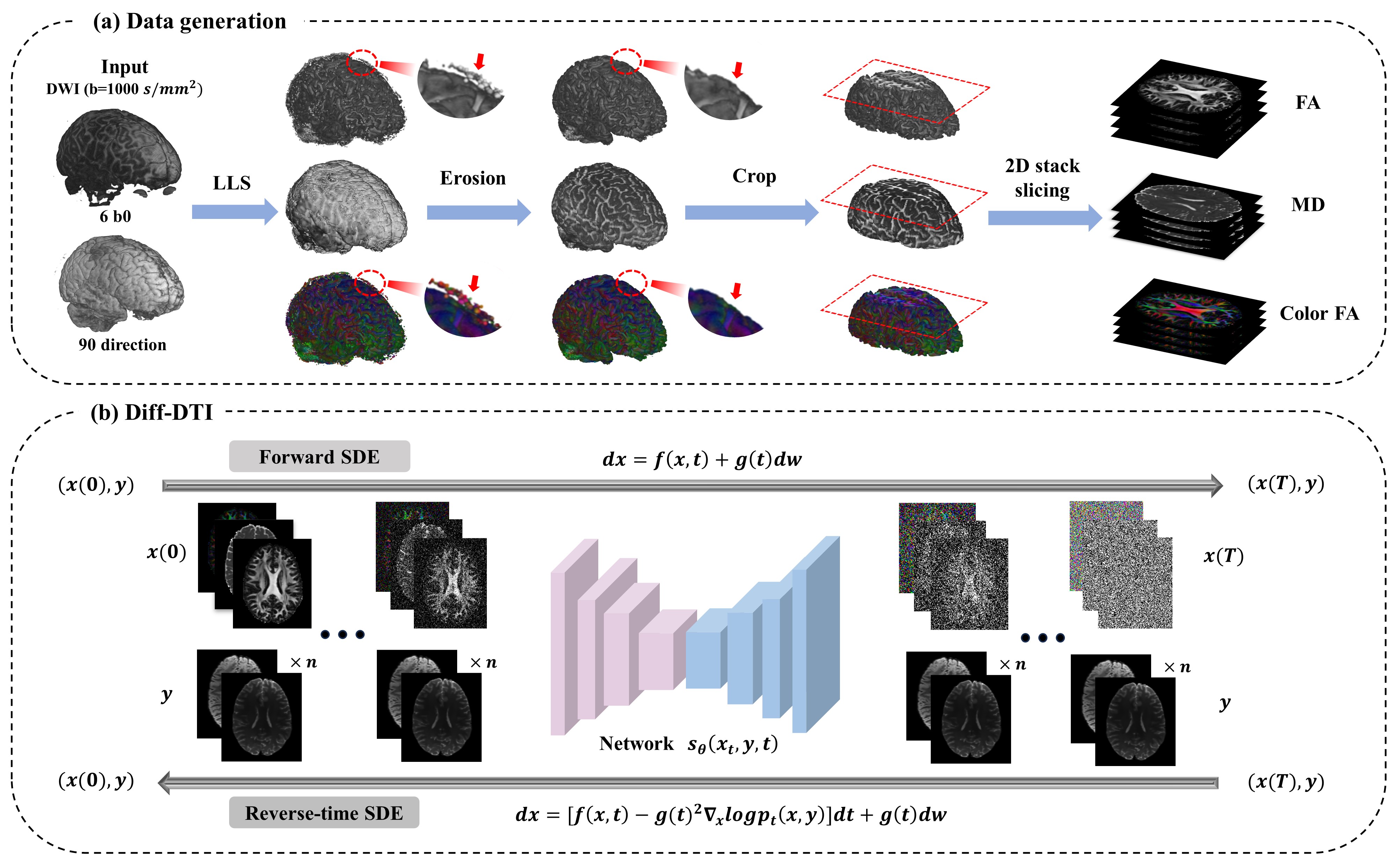}}
\caption{Overview of the processing flow of the Diff-DTI framework. (a) shows the processing flow of DTI maps (FA, MD, and Color FA). (b) shows Diff-DTI about overall process of the joint diffusion. }
\label{fig1}
\end{figure*}

\subsection{Joint Diffusion Model of n-directional DWIs and DTI map}
A joint diffusion model is constructed to generate DTI maps from \(n\)-directional DWIs. Initially, we concatenate the DTI map \((x)\) with DWIs \((y)\) along the channel dimension to form multi-channel data. The forward diffusion process is characterized by progressively adding noise to the original DTI map \(x(0)\) over T time steps, while y is fixed. This diffusion process transitions the original data distribution \(x(0)\sim p_{0}(x)\) to a Gaussian distribution \(x(T)\sim p_{T}(x)\), and \(x(t)\sim p_{t}(x)\) is intermediate distribution between \(x(0)\) and \(x(T)\), \(t\in[0, \mathrm{~T}]\), where \(x(T)\) is the outcome by adding noise to \(x(0)\) gradually based on the forward SDE. The forward SDE is given as:
    \begin{equation}\label{eqn-1} 
    dx=f(x,t)dt+g(t)dw
    \end{equation}
where \(f(x,t)\) and \(g(t)\) are the drift and diffusion coefficients of the SDE, \(dt\) is a negative infinitesimal time step, \(w\) represents the standard Wiener process. Specifically, we adopt the variance explosion VE-SDE mentioned in reference\cite{song2020improved}, setting \(f(x,t)=0\) and \(g(t)=\sqrt{\frac{d[\sigma^2(t)]}{dt}}\), \(\sigma(t)\) is the given coefficient to control the noise level. The forward SDE can be simplified as:
    \begin{equation}\label{eqn-2} 
    dx=\sqrt{\frac{d[\sigma^2(t)]}{dt}} dw     
    \end{equation}
The reverse process aims to perform the opposite transformation, i.e. reconstruct the DTI map \(x(0)\) from the noisy data under the condition \(y\). The reverse SDE is given as:
    \begin{equation}\label{eqn-3} 
    dx=[f(x,t)-g(t)^2\nabla_xlogp_t(x|y)]dt+g(t)dw.   
    \end{equation}
According to the properties of logarithm and derivative
    \begin{equation}\label{eqn-4} 
    \begin{aligned}
    \nabla_{x}logp_{t}(x|y)&=\nabla_{x}\big(logp_{t}(x,y)-logp_{t}(y)\big)\\
    &=\nabla_{x}logp_{t}(x,y)\end{aligned}
    \end{equation}
The reverse-time SDE can be rewritten as:
    \begin{equation}\label{eqn-4} 
    dx=[f(x,t)-g(t)^2\nabla_xlogp_t(x,y)]dt+g(t)dw 
    \end{equation}
where \(w\) is the Wiener process in the reverse process, \(p_t (x,y)\)  is a prior distribution that is known and close to the Gaussian distribution.  \(\nabla_xlogp_t(x,y)\) is the score function of \(p_t (x,y)\) and is an unknown quantity. Here, the score function of the joint distribution \(p_t (x,y)\) is approximated by establishing a neural network \(s_{\theta}(x_{t},y,t)\) using the score-matching method\cite{song2020improved}, i.e. \(s_\theta(x_t,y,t)\approx\nabla_xlogp_t(x,y)\). By learning this score function, the diffusion process can be reversed to generate the DTI map of the given DWIs using the reverse SDE as follows:
    \begin{equation}\label{eqn-4} 
    \begin{aligned}
    dx &= [f(x,t)- g(t)^2s_\theta(x_t, y, t)]\,dt+g(t)\,dw \\
       &= -\frac{d[\sigma^2(t)]}{dt}s_\theta(x_t, y,t)\,dt + \sqrt{\frac{d[\sigma^2(t)]}{dt}} \, dw
    \end{aligned}
    \end{equation}

%% fig2
\begin{figure*}[!htp]
\centerline{\includegraphics[width=0.9\textwidth]{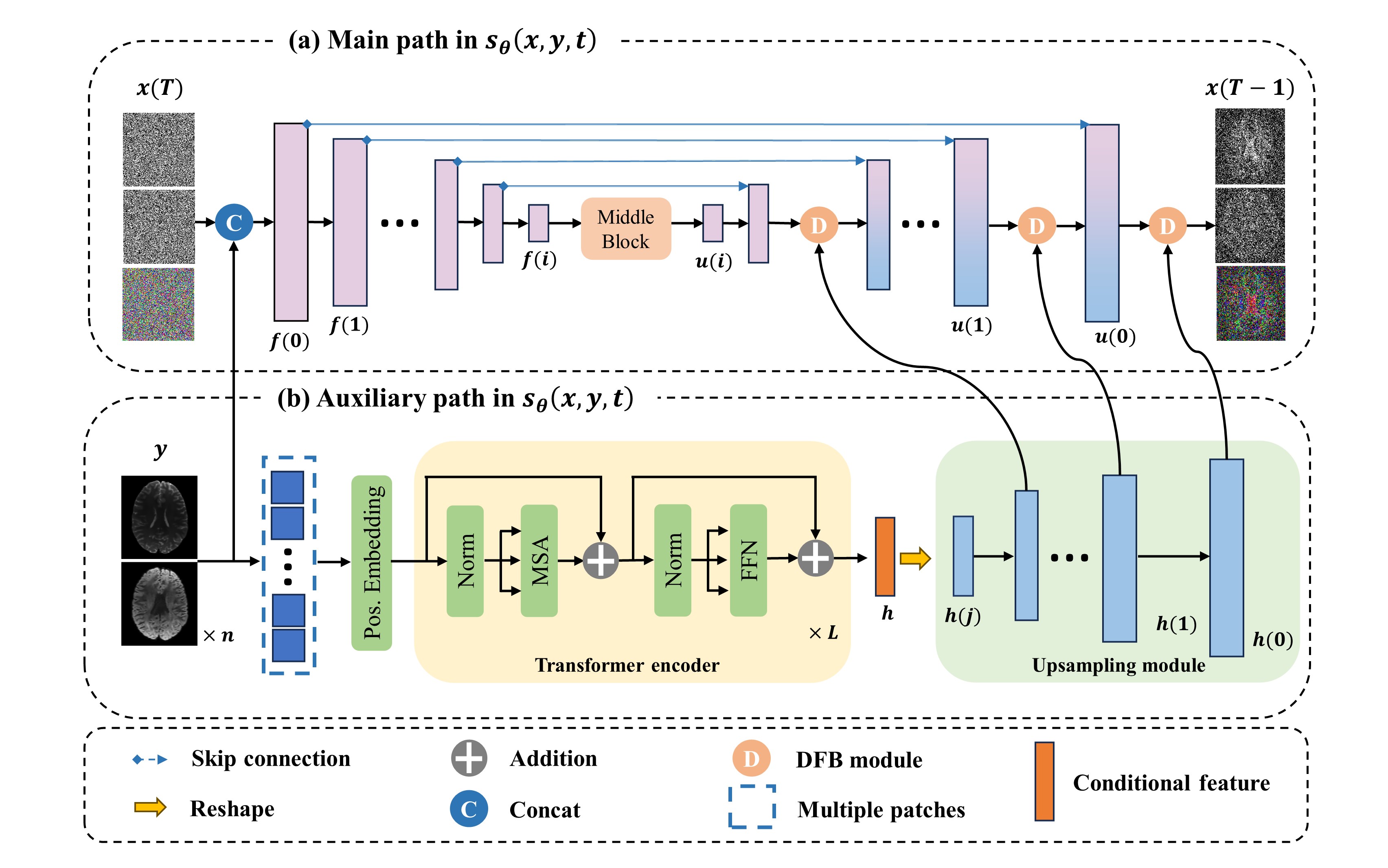}}
\caption{Overview of the network structure \(s_{\theta}(x_{t},y,t)\) in Diff-DTI framework. (a) shows the main path codec of the network. (b) shows auxiliary path for feature enhancement fusion.}
\label{fig2}
\end{figure*}
    
The training process of \(s_{\theta}(x_{t},y,t)\) is as follows. First, Gaussian noise is added to the DTI map using the forward process of (2), and then concatenated with the \(n\)-directional DWIs to obtain the input of \(s_{\theta}\). The added Gaussian noise has a zero mean, and its variance varies with \(t\), which is determined by the conditional probability density function \(p_{0t}\big(x(t)\big|x(0)\big)=N(x(t);x(0),\delta(t)^{2}\). Then, in reverse process, we train \(s_{\theta}(x_{t},y,t)\) to approximate the logarithmic gradient of the conditional probability density function \(\nabla_x\log p_{0t}\big(x(t)\big|x(0)\big)=- \frac{x(t)-x(0)}{\delta^2(t)}\), thus efficiently guiding \(x(t)\) towards \(x(0)\) under condition \(y\). The optimization objective is:
    \begin{equation}\label{eqn-4} 
    L=E_{x(0)}E_{x(t)|x(0)}\left[\lambda(t)\left|\left|s_{\theta}(x(t),y,t)+\frac{x(t)-x(0)}{\delta^{2}(t)}\right]\right||_{2}^{2}\right]
    \end{equation}
where \(\lambda(t)\) is a time-dependent weighting function.

\subsection{Network architecture with FEFM}
A novel network architecture is proposed for Diff-DTI, aimed at enhancing the global feature extraction capability of \(s_{\theta}\), thereby enabling more accurate capture of the score function. The network consists of the main and auxiliary paths, as illustrated in Fig. 2. The main path employs a U-Net architecture similar to that used by Song\cite{song2020improved}. This U-Net model (shown in Fig. 2(a)) encodes the noisy image \(x\) into multi-scale feature maps \(f(i),i=1,\cdots,S\), and \(S\) is the depth of the U-Net. The auxiliary pathway integrates the FEFM, which consists of a Feature Enhancement Network (FEN) parallel to the U-Net and a series of multiscale fusion layers consisting of DFBs. Among them, the FFN is used to extract global and high-frequency detailed features. The feature layers of different scales in the two pathways are fused by conditionally guided DFBs to improve the performance of Diff-DTI.

\subsubsection{Feature Enhancement Network}
The FEN aims to capture the global features from DWIs through self-attention mechanism. The architecture of FEN is shown in Fig. 2(b). The FEN is mainly constructed by two modules: Transformer encoder module and Upsampling module. The transformer encoder module has L layers for feature encoding. Each layer consists of two Normalizations (Norm), a Multi-Head Self-Attention (MSA) and a Feed-Forward Multilayer Perceptron (FFN). The MSA allows the model to focus on information across different layers and conditions, with each self-attention matrix being:
    \begin{equation}\label{eqn-4} 
    Z_m=Attention(Q_m,K_m,V_m)=softmax(\frac{Q_mK_m^T}{\sqrt{d_k}})\cdot V_m
    \end{equation}
    \begin{equation}\label{eqn-4} 
    Z=Concat(Z_1,...,Z_m)W_0
    \end{equation}
where \(Q, K,\) and \(V\) represent the query, key, and value vectors obtained by multiplying the \(n\)-directional DWIs image sequences with different weight matrices \(W_m^Q, W_m^K\) and \(W_m^V\), respectively. m refers to number of heads in MSA. \(d_k\) is the dimension of the key vector. Z is the concatenation of the outputs from each head \(Z_m\). \(W_0\) is the output weight matrix. The FFN, functioning as a multilayer perceptron, aims to extract key features. It consists of two fully connected layers (FC) that take the attention matrix \(Z\) as input, with its output features being:
    \begin{equation}\label{eqn-4} 
    f=FFN(Z)=\max(0,ZW_1+b_1) W_2+b_2
    \end{equation}
where \(W_1\) and \(W_2\) are the weight matrices, \(b_1\) and \(b_2\) are the bias terms. The upsampling model includes a LayerNorm2d layer normalization and a transposed convolution layer with the GeLU activation function, which obtains multi-scale feature layers with detailed information.

The pipeline of FEN is as follows. The input of FEN is the combination of one b0 and \(n\)-directional DWIs. To conform to the requirement of the transformer encoder, the input images are initially segmented into multiple patches and then appended with 
positional embeddings to create image sequences. These sequences are subsequently fed into the Transformer Encoder to establish the initial conditional feature \(h\). Then upsampling module is applied to \(h\) to extract conditional features at multiple scales, i.e. \(h(j),j=1,\cdots,S-2,\) through bilinear interpolation. \(h(j)\) is then fused with the multi-scale feature layers from the main path via the DFB module. The use of FEN directly provides encoded multi-scale features that can be fuses with feature layers from the main path, without the need for extra priors to model the latent representation.

\subsubsection{Diffusion Feature Fusion Block}
DFB aims to fuse the multi-scale feature maps of the main path and the auxiliary path. The architecture of DFB is illustrated in Fig. 3. \(f(i)\) and \(h(j)\) represents the feature maps extracted from the main and auxiliary pathways, respectively. \(h(j)\) is first processed through a self-attention layer followed by a convolutional layer to learn the adaptive mapping parameters. Two adaptive mapping parameters are learned, i.e., the scaling factor \(s_1 (i)\) and translation factor \(s_2 (i)\). Then \(f(i)\) is modulated by \(s_1 (i)\)  and \(s_2 (i)\) flexibly through multiplication and addition, respectively, resulting the enhanced and fused feature layer \(u(i)\). The formular of the process is defined as:
    \begin{equation}\label{eqn-4} 
    s_1(i)=Conv_\alpha(A(h(j)))
    \end{equation}
    \begin{equation}\label{eqn-4} 
    s_2(i)=Conv_\beta(A(h(j)))
    \end{equation}
    \begin{equation}\label{eqn-4} 
    u(i)=f(i)\cdot s_{1}(i)+s_{2}(i)
    \end{equation}
where \(A(\cdot)\) represents the self-attention layer. \(Conv(\cdot)\) represents the convolution layer with ReLU activation function, \(\alpha\) and \(\beta\) are the weight parameters of the convolution layer. DFB utilizes the built-in self-attention layers to pay an attention key information and adapts the scaling factors to make targeted adjustments to the fused features. Relative to the concatenation fusion method, DFB significantly improves the capability of information retention, and is therefore particularly suitable for scenarios that require precise spatial adjustment based on specific condition information.

%% fig3
\begin{figure}[!htbp]
\centering
\includegraphics[width=0.95\columnwidth]{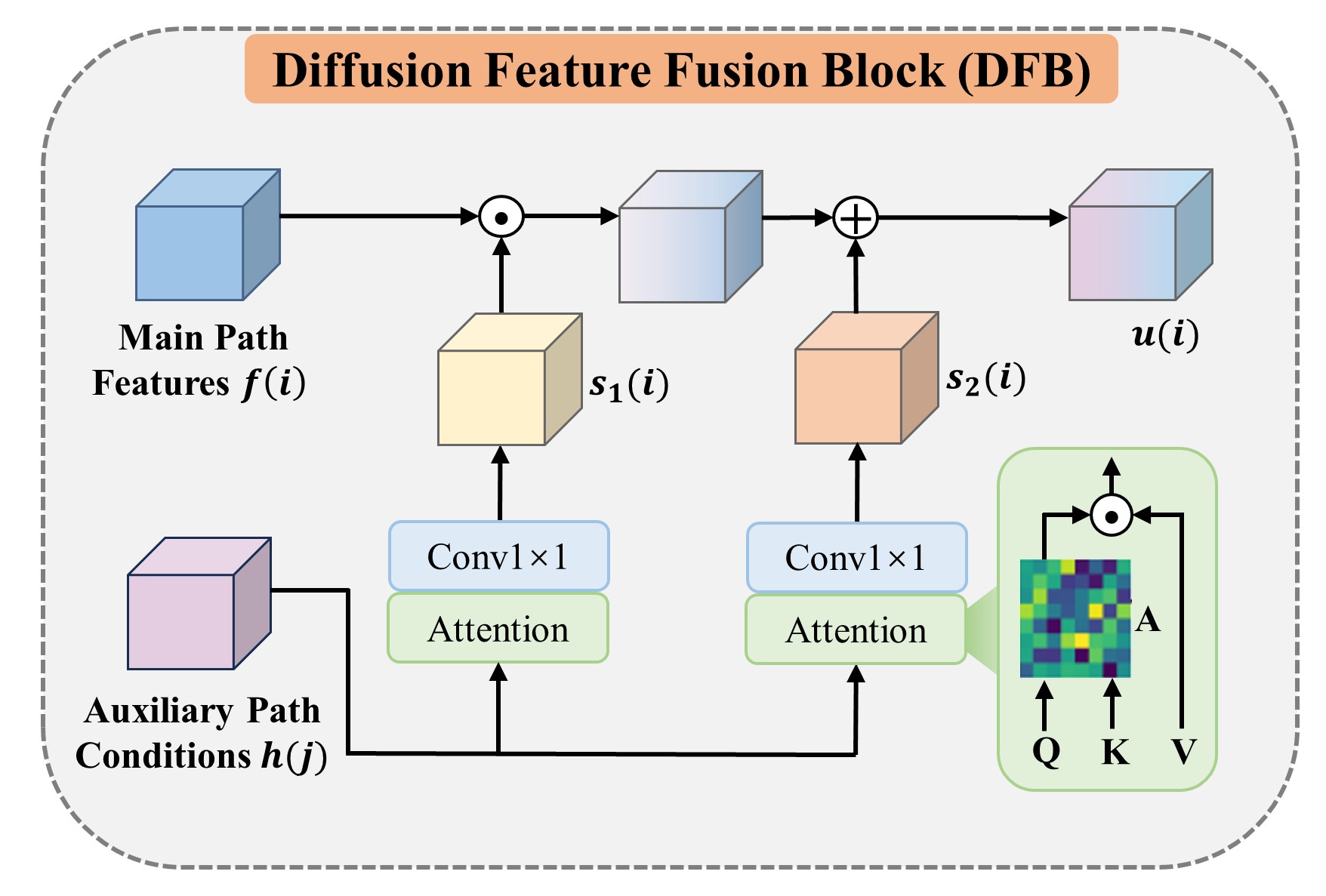}
\caption{Overview of the fusion of DFB the main and auxiliary path feature layers, where two adaptive mapping parameters \(s_1 (i)\) and \(s_2 (i)\) are used to guide the fusion of \(f(i)\) and \(h(j)\). }
\label{fig3}
\end{figure}

%% fig4
\begin{figure*}[!htp]
\centerline{\includegraphics[width=\textwidth]{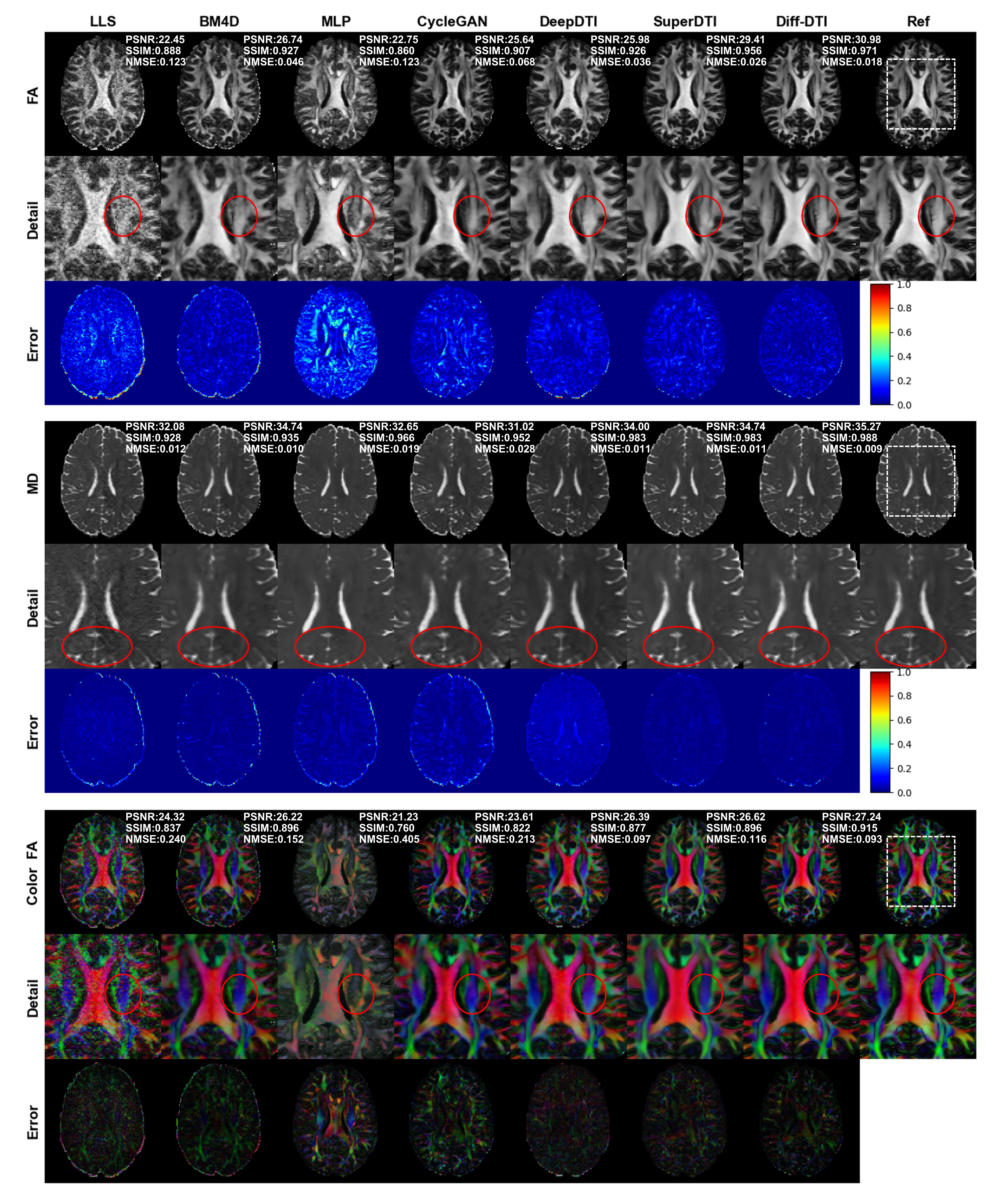}}
\caption{A comparison of FA, MD, and Color FA generated from one b0 and six DWIs. The first line of each map represents a single slice generated by different methods, and the three metrics of the slice. The ‘Detail’ row shows a partially enlarged area within the white dashed rectangular box of the Ref images for closer inspection, while the "Error" row presents their error maps. The PSNR, SSIM, NMSE relative to the references were given.}
\label{fig4}
\end{figure*}

\section{Experiments}
\subsection{Datasets}
The DWI data used in this study are from the Human Connectome Project (HCP) Young Adult dataset, obtained using a multiband diffusion sequence on a 3T scanner (Skyra, Siemens Healthcare)\cite{elam2021human}. The dataset was collected using three different gradient tables, uniformly distributed over three q-space shells at b = 1000, 2000, and 3000 \(s/mm^2\). Each shell comprises 90 diffusion-gradient directions and six b0 images. Other imaging parameters were listed in Table 1. The DWI images with b = 1000 \(s/mm^2\) were selected in this study, involving 203 subjects. The middle 50 slices of each subject, which have regular brain shapes, were included in this study. All data underwent preprocessing through the HCP-provided pipelines, including corrections for EPI distortion, eddy-current-induced distortion, and motion shift. The dataset was divided into a training subset (183 subjects, 9150 slices) and a testing subset (20 subjects, 1000 slices). Image was normalized and image size was zero padded from (145,174) to (192,192) to facilitate the network operations. The reference DTI maps were calculated using the Linear Least Squares fitting (LLS) from 90 DWIs and 6 b0, serving as the label for training and gold standard for testing. A morphological erosion procedure was applied to the reference DTI maps to remove singular points derived from noise at the brain boundary.

% TABLE1
\begin{table}[htbp]
\centering
\caption{The imaging parameters of the HCP DWI dataset}
\label{table}
\renewcommand{\arraystretch}{1.2} % 调整行间距
\setlength{\tabcolsep}{25pt}
\begin{tabular}{cc} 
\hline
\multirow{2}{*}{Parameter} & Public Dataset \\ \cline{2-2} 
                           & HCP Dataset    \\ \hline
TR & 5520 ms \\
TE & 89.5 ms \\
Flip Angle & 78 deg \\
Refocusing flip angle FOV & 160 deg \\
FOV & 180×210(PE×RO) \\
Matrix & 145×174(PE×RO) \\
Slice thickness & 1.25mm \\
Multiband factor & 3 \\
Echo spacing & 0.78ms \\
BW & 1488 HZ/Px \\
Phase partial Fourier & 6/8 \\ \hline
\end{tabular}
\label{tab1}
\end{table}

\subsection{Experiment setup}
The Diff-DTI method was compared with traditional LLS\cite{jiang2006dtistudio}, Block-Matching and 4D Filtering (BM4D) methods\cite{maggioni2012nonlocal}, as well as the deep-learning based methods including Multilayer Perceptron (MLP)\cite{ronneberger2015u}, CycleGAN\cite{zhu2017unpaired}, DeepDTI\cite{tian2020deepdti}, and SuperDTI\cite{li2021superdti}. For Diff-DTI, we used the Adam optimizer at a learning rate of  \(2\times10^{-4}\), with a maximum gradient clipping value of 1.0 and a 0.999 exponential moving average (EMA) method for model parameters. The noise level control utilized noise scales \(\sigma_{max}=348\) and \(\sigma_{min}=0.01\). During the inference phase, the iteration counts \(N=1000\), each iteration including steps for the predictor and corrector\cite{song2020improved}. The reconstruction process for a single DTI map took about 2 minutes on the GPU. For the comparison methods, MLP was a 4-layer multi-head MLP network trained for 100 epochs. CycleGAN was trained for 200 epochs. DeepDTI took DWIs as input and outputted high-quality DWIs, and then used LLS to generate maps; it was trained for 100 epochs. SuperDTI was an 8-layer convolution and deconvolution network, which took a 21×21 patch as input and was trained to 100 epochs. These methods used MSE as the loss function and utilized the Adam optimizer with a learning rate of \(2\times10^{-4}\); other settings were consistent with their original texts. All experiments were conducted on PyTorch and an NVIDIA RTX A100 GPU, with a batch size of 8, and all performed the same data processing, including morphological erosion.

\subsection{Evaluation Metrics}
The performance of Diff-DTI was assessed through three metrics: peak signal-to-noise ratio (PSNR), structural similarity index (SSIM), and quantization error (NMSE). The formula of PSNR is:
    \begin{equation}\label{eqn-4} 
    PSNR=20\cdot log_{10}\left(\frac{MAX_{ref}}{\sqrt{MSE}}\right)
    \end{equation}
where \(MAX_{ref}\) represents the maximum intensity of the reference DTI map, and MSE is the mean square error between the reference and generated DTI map. SSIM measures the similarity between the reference and generated DTI maps with the value range of 0 to 1. The formula of SSIM is: 
    \begin{equation}\label{eqn-4} 
    SSIM=\frac{\bigl(2\mu_{ref}\mu_{rec}+c_1\bigr)\bigl(2\sigma_{ref,rec}+c_2\bigr)}{\bigl(\mu_{ref}^2+\mu_{rec}^2+c_1\bigr)\bigl(\sigma_{ref}^2+\sigma_{rec}^2+c_1\bigr)} 
   \end{equation}
where \(\mu_{ref}\)  and \(\mu_{rec}\) represent the mean luminance of the reference and generated DTI map, respectively, \(\sigma_{ref}^2\) and \(\sigma_{rec}^2\) are their corresponding variances. \(\sigma_{ref,rec}\) is the covariance of the reference and generated DTI map.  \(c_1\) and \(c_2\) are small constants added to stabilize the division. NMSE is used to measure the difference between the reference and generated DTI map, with smaller values indicating better performance. The formula of NMSE is:
    \begin{equation}\label{eqn-4} 
    NMSE=\frac{\sum_{i=1}^N(ref_i-rec_i)^2}{\sum_{i=1}^Nref_i^2}
    \end{equation}
where \(ref_i\) and \(rec_i\) represent the pixel values of the reference and generated DTI map.

\section{Results}
\subsection{Generation of DTI maps}
Fig. 4 shows the maps of FA, MD, and Color FA generated by different methods using one b0 and six DWIs images. The DTI maps generated by Diff-DTI is visually comparable to the reference maps and surpass other methods in quantitative metrics. The method of LLS, BM4D, and DeepDTI exhibit higher noise level than deep learning-based method, particularly in edge regions. This observation is consistent with the notion that networks are effective in noise suppression. As indicated by the red circles, Diff-DTI can accurately reconstruct complex textural details. This is especially evident in the characteristic striations induced by the grey matter bridges in FA and Color FA maps, addressing the issues of excessive smoothing and loss of detail observed in other methods.

Table 2 presents the average quantitative metrics on the test subset using different methods. Diff-DTI achieves the best performance in all metrics for FA and MD. In Color FA, Diff-DTI demonstrates the best SSIM and NMSE among all the comparison methods, but with slightly lower PSNR than DeepDTI.

Fig. 5 displays the DTI maps generated by MLP, CycleGAN, SuperDTI, and Diff-DTI using one b0 and three DWIs images. The minimum number of DWIs required by LLS, BM4D, and DeepDTI methods is 6, which is larger than the used DWIs in this experiment. Therefore, these three methods were excluded from comparison. Generally speaking, the image quality of all compared methods slightly degrade relative to Fig. 4 due to the reduction of DWI number. We also used red circles to highlight the detail structures of DTI maps in Fig. 5. Diff-DTI still achieves the best visually quality among the compared methods for FA, MD, and Color FA maps, although the FA map is slightly blurry compared with using 6 DWIs. In contrast, the DTI maps of MLP exhibit severe artifacts. The results of CycleGAN and SuperDTI are even smoother compared with Diff-DTI.

Table 3 presents the average quantitative metrics on the test subset with one b0 and three DWI images. Diff-DTI achieved the best results across all indices for FA, MD, and Color FA. Compared to the results in Table 2, the quantitative metrics of Diff-DTI are slightly decrease due to the reduction in DWIs. Note that the change in SSIM is very limited, which suggests that Diff-DTI can preserve fine structures in DTI maps with fewer DWIs. Other methods, especially MLP and SuperDTI, exhibit obvious decline in the metrics of FA and Color FA.

%% fig5
\begin{figure}[!htp]
\includegraphics[width=\columnwidth]{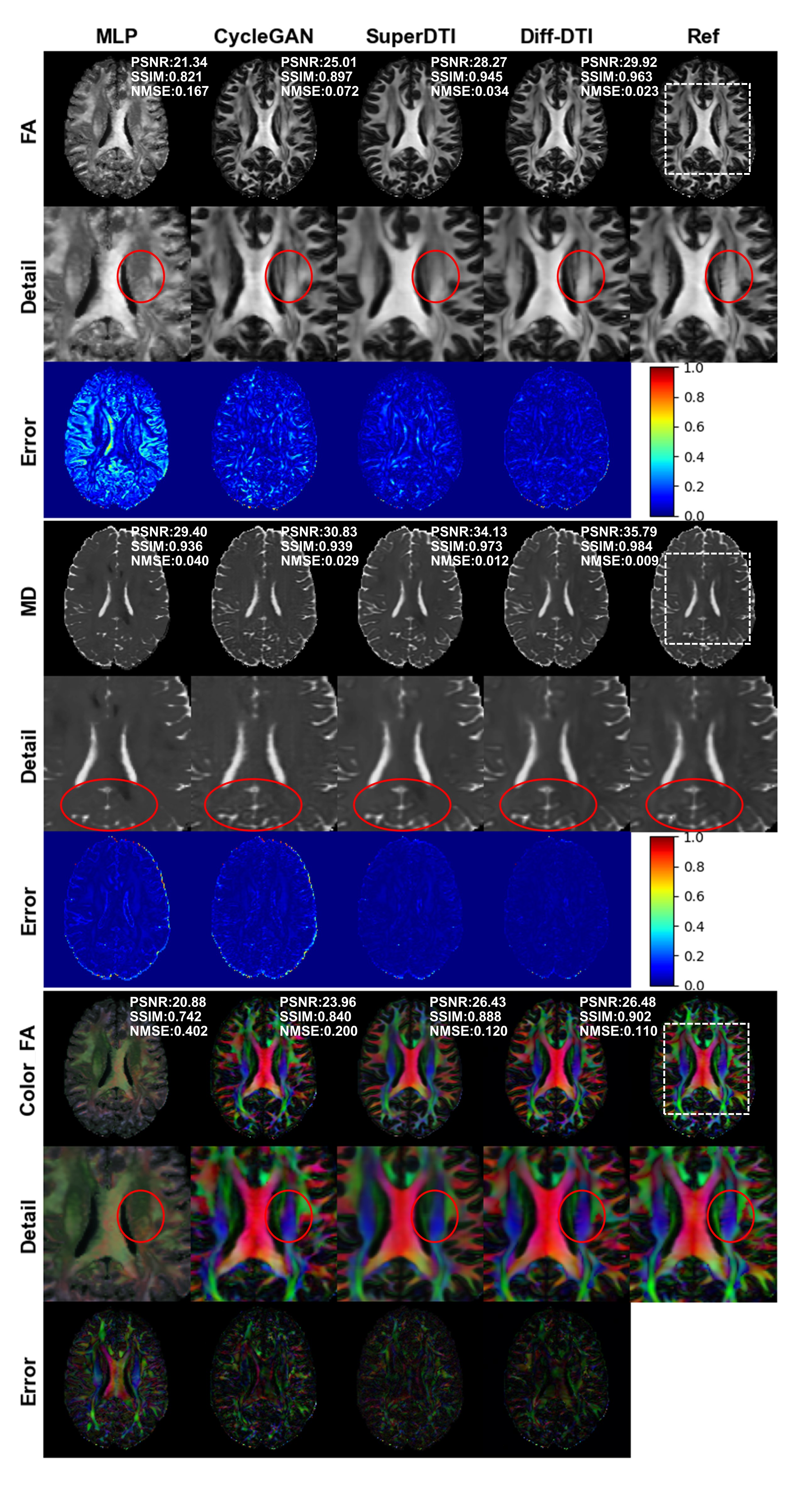}
\caption{A comparison of FA, MD, and Color FA generated from one b0 and three DWIs. The first line of each map represents a single slice generated by different methods, and the three metrics of the slice. The ‘Detail’ row shows a partially enlarged area within the white dashed rectangular box of the Ref images for closer inspection, while the "Error" row presents their error maps. The PSNR, SSIM, NMSE relative to the references were given.}
\label{fig5}
\end{figure}

% TABLE2
\begin{table}[htbp]
\centering
\caption{Quantitative assessment of FA, MD, and Color FA using different methods on one b0 and six DWIs.}
\label{table}
\renewcommand{\arraystretch}{1.5}
\setlength{\tabcolsep}{1pt}
\begin{tabular}{ccccccclll}
\hline
\multirow{2}{*}{Methods} & \multicolumn{3}{c}{FA} & \multicolumn{3}{c}{MD} & \multicolumn{3}{c}{Color FA} \\ \cline{2-10} 
 & PSNR & SSIM & NMSE & PSNR & SSIM & NMSE & PSNR & SSIM & NMSE \\ \hline
LLS & 20.29 & 0.842 & 0.122 & 36.00 & 0.928 & 0.075 & 22.54 & 0.803 & 0.240 \\
BM4D & 23.81 & 0.874 & 0.118 & 36.60 & 0.930 & 0.035 & 25.76 & 0.843 & 0.172 \\
MLP & 25.66 & 0.913 & 0.107 & 34.29 & 0.976 & 0.019 & 24.12 & 0.831 & 0.361 \\
CycleGAN & 25.85 & 0.909 & 0.104 & 31.71 & 0.960 & 0.028 & 25.26 & 0.858 & 0.293 \\
DeepDTI & 27.83 & 0.949 & 0.044 & 37.33 & 0.987 & 0.010 & \textbf{29.32} & 0.915 & 0.123 \\
SuperDTI & 30.31 & 0.963 & 0.036 & 36.14 & 0.987 & 0.011 & 29.02 & 0.912 & 0.154 \\
Diff-DTI & \textbf{31.39} & \textbf{0.972} & \textbf{0.028} & \textbf{37.76} & \textbf{0.991} & \textbf{0.008} & 29.06 & \textbf{0.930} & \textbf{0.121} \\ \hline
\end{tabular}
\label{tab1}
\end{table}

% TABLE3
\begin{table}[htbp]
\centering
\caption{Quantitative assessment of FA, MD, and Color FA using different methods under one b0 and three DWIs conditions.}
\label{table}
\renewcommand{\arraystretch}{1.5}
\setlength{\tabcolsep}{1pt}
\begin{tabular}{llllllllll}
\hline
\multicolumn{1}{c}{\multirow{2}{*}{Methods}} & \multicolumn{3}{c}{FA} & \multicolumn{3}{c}{MD} & \multicolumn{3}{c}{Color FA} \\ \cline{2-10} 
\multicolumn{1}{c}{} & \multicolumn{1}{c}{PSNR} & \multicolumn{1}{c}{SSIM} & \multicolumn{1}{c}{NMSE} & \multicolumn{1}{c}{PSNR} & \multicolumn{1}{c}{SSIM} & \multicolumn{1}{c}{NMSE} & PSNR & SSIM & NMSE \\ \hline
MLP & 23.66 & 0.875 & 0.167 & 33.65 & 0.968 & 0.019 & 23.59 & 0.812 & 0.400 \\
CycleGAN & 25.55 & 0.905 & 0.112 & 30.92 & 0.954 & 0.035 & 25.08 & 0.857 & 0.309 \\
SuperDTI & 29.45 & 0.954 & 0.044 & 35.47 & 0.982 & 0.012 & 28.43 & 0.906 & 0.168 \\
Diff-DTI & \textbf{30.55} & \textbf{0.967} & \textbf{0.035} & \textbf{37.07} & \textbf{0.991} & \textbf{0.009} & \textbf{28.47} & \textbf{0.923} & \textbf{0.139} \\ \hline
\end{tabular}
\label{tab1}
\end{table}

%% fig6
\begin{figure}[!htp]
\centering
\includegraphics[width=\columnwidth]{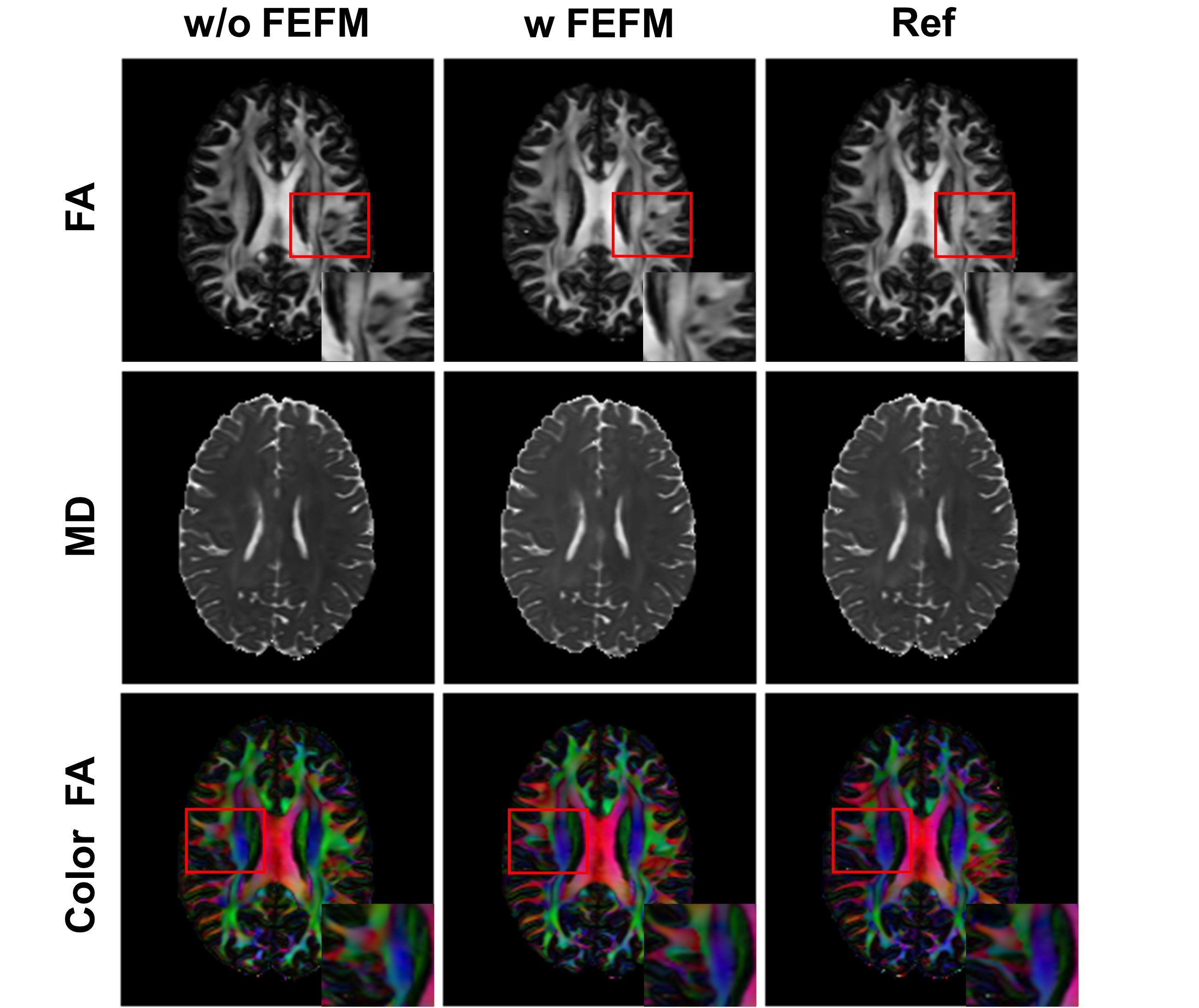}
\caption{Effect of FEFM on Diff-DTI generated FA, MD, and Color FA. The first column shows FEFM removed, the second column shows FEFM added, and the third column shows the reference image. The lower right corner shows the local magnified area in the red rectangular box.}
\label{fig6}
\end{figure}

% TABLE4
\begin{table}[htbp]
\centering
\caption{Quantitative assessments of FA, MD, and Color FA were conducted under conditions of one b0 and three DWIs to evaluate the impact of the FEFM.}
\label{table}
\renewcommand{\arraystretch}{1.5}
\setlength{\tabcolsep}{1pt}
\begin{tabular}{llllllllll}
\hline
\multirow{2}{*}{Diff-DTI} & \multicolumn{3}{c}{FA} & \multicolumn{3}{c}{MD} & \multicolumn{3}{c}{Color FA} \\ \cline{2-10} 
 & \multicolumn{1}{c}{PSNR} & \multicolumn{1}{c}{SSIM} & \multicolumn{1}{c}{NMSE} & \multicolumn{1}{c}{PSNR} & \multicolumn{1}{c}{SSIM} & \multicolumn{1}{c}{NMSE} & PSNR & SSIM & NMSE \\ \hline
\begin{tabular}[c]{@{}c@{}}w/o \\ FEFM\end{tabular} & 30.45 & 0.964 & 0.035 & 37.38 & 0.990 & 0.009 & 28.14 & 0.907 & 0.132 \\
\begin{tabular}[c]{@{}c@{}}w \\ FEFM\end{tabular} & \textbf{31.39} & \textbf{0.972} & \textbf{0.028} & \textbf{37.76} & \textbf{0.991} & \textbf{0.008} & \textbf{29.06} & \textbf{0.930} & \textbf{0.121} \\ \hline
\end{tabular}
\label{tab1}
\end{table}

%% fig7
\begin{figure*}[!htp]
\centerline{\includegraphics[width=0.9\textwidth]{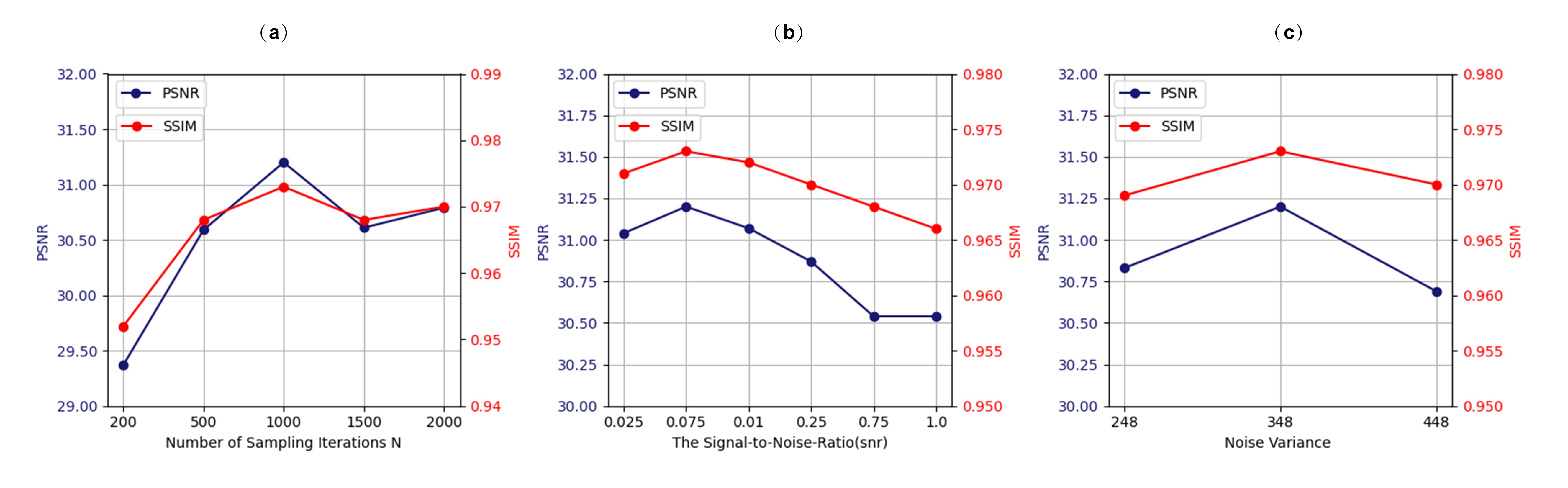}}
\caption{Validation of PSNR and SSIM curves for different hyperparameters on a FA generation task with one b0 and six DWIs. (a) shows the effect of the number of sampling iterations (N); (b) shows the effect of SNR on the sampling process; and (c) shows the effect of noise variance.}
\label{fig}
\end{figure*}

\subsection{Ablation Study}
We performed an ablation experiment to demonstrate the effectiveness of FEFM in Diff-DTI. Table 4 presents the results of Diff-DTI with and without FEFM. The removal of FEFM results in decreased PSNR, SSIM, and NMSE, particularly in the generation of FA and Color FA. To better show the differences in detail reconstruction with and without FEFM, we enlarged the area of more contrasting FA and Color FA (indicated by the red rectangle in Fig. 6). As shown in Figure 6, the w/o FEFM Diff-DTI exhibits irregular shapes and artifacts for the generation of fine pore-like structures in FA, and Color FA shows obvious chromatic aberration and discontinuous texture details. The absence of the FEFM led to failure of Diff-DTI to reconstruct high-fidelity details, while w FEFM Diff-DTI achieves similar results to Ref, which demonstrating the FEFM crucial role in effectively extracting high-frequency detail information.	

We performed an ablation experiment on three key hyperparameters that impact the performance of Diff-DTI. The analysis was performed on an FA generation using one b0 and six DWIs conditions. Firstly, we explored the number of sampling iterations (N), which directly influences the quality and efficiency of Diff-DTI. As shown in Fig. 7(a), by comparing five different settings of N= {200, 500, 1000, 1500, 2000}, we found that Diff-DTI achieves optimal performance and efficiency at N=1000. Next, we examined the effect of the “signal-to-noise ratio” (SNR) on noise management during the sampling process. As illustrated in Fig. 7(b), by adjusting several values of SNR, we determined that setting SNR= 0.075 provides the most effective balance between noise and artifact mitigation. Lastly, we investigated the impact of noise variance on the degree of perturbation each diffusion step imparts on the data, defined as \(\sigma(t)=\sigma_{min}(\frac{\sigma_{max}}{\sigma_{min}})^{t}\). Following Song's VESDE configuration, setting \(\sigma_{max}= 0.01\) and comparing various \(\sigma_{max}= \{248, 348, 448\}\).  As illustrated in Fig. 7(c), we observed as depicted that setting \(\sigma_{max}= 348\) a more stable diffusion and sampling process.

\section{Discussion}
In this study, we developed a novel model based on diffusion model, which accurately predicts DTI maps from a limited number of DWIs. Diff-DTI achieved accurate conditional bootstrap generation by inscribing the joint probability distribution modelling of DWIs and DTI maps. The model integrates FEFM, comprised of a Transformer-based FEN and a series of DFBs for multi-scale fusion, effectively capturing detailed information. Compared to other DTI fast imaging methods, Diff-DTI demonstrates higher reliability, breaking previous constraints that required a minimum of six DWI images to generate DTI maps. It not only accurately generates DTI maps but also provides finer detail representation. In addition, Diff-DTI also shows optimal results in various evaluation metrics.

Diff-DTI significantly increases the speed of DTI imaging. Conventional LLS require at least 30 DWIs in uniformly distributed orientations to meet clinical requirements and take about 5 - 10 minutes\cite{jones2013white}. In contrast, deep learning-based methods have reduced the need for DWIs to 6, shortening scanning time to less than 1 minute. Nevertheless, few studies have further reduced the number of DWIs to less than 6. As shown in Fig. 5 and Table 3, Diff-DTI breaks through this limitation by generating high-quality DTI maps with only 3 DWIs, further halving the scanning time. In addition, Diff-DTI is insensitive to the diffusion gradient direction of DWIs, thus simplifying the data preparation process. This contrasts with methods that require careful selection of uniformly distributed DWIs, e.g. DeepDTI obtains uniformly distributed DWIs by minimizing the condition number of the diffusion tensor transformation matrix\cite{tian2020deepdti}, while SuperDTI and MSIS-DTI use a uniform diffusion coding scheme\cite{li2021superdti,wang2024ultrafast}. Even with the use of three DWIs that are not uniformly distributed as input, Diff-DTI efficiently captures the data distribution of DTI maps and generate high-fidelity DTI maps comparable to the reference.

Diffusion models possess robust capabilities for modeling data distributions, making them highly effective in image-to-image translation tasks. It not only learns the mapping relationship between conditional input DWIs and output DTI maps, but also captures the distribution characteristics of DTI maps, which previous methods failed to achieve. Additionally, our joint diffusion model (Diff-DTI) can generate DTI maps that closely match reality without the need for additional training of classifiers or complex guided sampling. As illustrated in Fig. 4 and 5, the DTI maps generated by Diff-DTI are more realistic compared to other methods because it avoids the homogenization effect often seen with other techniques\cite{loizillon2024leveraging,loizillon2024detecting,robertson2018effects}. This effect is usually not expected to occur as it can make the pathological features of lesions or injuries on DTI maps appear similar or blurred across different patients, potentially obscuring critical pathological information and hindering effective diagnosis by physicians.

At the same time, we observed that deep learning-based methods underperform in predicting directional parameters (e.g., Color FA) compared to non-directional parameters (e.g., FA and MD). This disparity mainly stems from the multi-channel complexity and high dimensionality of Color FA, which complicates model training and often leads to the "curse of dimensionality," thereby reducing performance\cite{youm2016image,souza2020dual}. Additionally, Color FA is a map that includes both diffusion intensity and directional information, where the continuity of directions and small mutations reflect critical variations in biological tissues that are difficult to accurately capture by deep learning methods. Despite these challenges, as shown in the results section, Diff-DTI has made significant progress in processing Color FA, outperforming other methods in both visualization and performance metrics. This is due to Diff-DTI's robust data distribution modeling capability and its architectural inclusion of a Transformer, which effectively handles long-range dependencies through its self-attention mechanism, enhancing the capture of directional information.

The performance superiority of Diff-DTI primarily stems from its FEFM. The FEFM optimizes the generation of DTI maps by incorporating both global and local information, effectively restoring complex details. In contrast, existing methods typically employ CNNs or Transformers combined with an L2 loss function, which results in overly smooth DTI maps with lost details\cite{lyu2022conversion}. Although some studies have attempted to improve the loss function by introducing regularizes based on singular value decomposition to reduce detail loss, they still cannot overcome the homogeneity problem\cite{li2024dimond, chen2022image}. Additionally, traditional diffusion models primarily utilize a single UNet or a modified UNet as the score function or denoising network\cite{luo2023refusion,kim2024adaptive,zhang2024dosediff}. For instance, in Reformer's denoising network design, Nonlinear Activation-Free Blocks (NAFBlocks) replace the convolutional blocks of UNet and are activated through SimpleGate, directly processing features to preserve more original information. While this simplified network structure design can improve gradient flow, it limits the capacity of model to handle complex data distributions, especially for learning structurally complex data such as DTI maps\cite{luo2023refusion}. In view of this, we designed an auxiliary path containing a Transformer-based FEN, extracting features from DWIs and perceiving global information. Diff-DTI also integrates a DFB for multi-scale feature fusion to achieve an effective fusion global and local information, using global insights to guide local detail restoration and prevent excessive smoothing and homogenization. As shown in Fig. 6 and Table 4, the inclusion of the FEFM allows Diff-DTI to produce more realistic detail generation and improves the handling of color direction in Color FA.

This study marks the first application of diffusion models in the field of DTI imaging, and it represents a preliminary exploration. The results demonstrate that the generated DTI maps are encouraging in terms of accuracy and detail. However, the research has several limitations: First, limited by the size of the dataset, we have not yet evaluated the performance of the method on different MRI scanners. Secondly, Diff-DTI relies on Markov chain generation, which results in relatively slow generation speeds. For this reason, future work will explore efficient training and sampling algorithms, and consider introducing a physical model of DTI to help capture the physical attributes and biological tissue features in DTI maps. In the meantime, we plan to continue collecting DTI data from various scanners for enable extensive exploration and validation.

\section{Conclusion}
We proposed Diff-DTI, a feature-enhanced joint diffusion model for accelerated DTI imaging. Diff-DTI uses six or even fewer DWIs as conditions for conditionally generative modelling of the distribution of DTI maps by means of a joint diffusion model and a FEFM. Compared with other DTI imaging methods, Diff-DTI generated reliable DTI maps with fine details and achieved SOTA performance on the HCP dataset.

\section*{ACKNOWLEDGEMENT}
This work was supported in part by the National Natural Science Foundation of China under Grant 63232119, 12226008, 62125111, U1805261, 62106252, 62206273, and 62201561, in part by the National Key R\&D Program of China under Grant 2020YFA0712200 and Grant 2021YFF0501402, in part by the Guangdong Basic and Applied Basic Research Foundation under Grant 2023B1212060052, and in part by the Shenzhen Science and Technology Program under Grant RCYX2021060910444089 and JCYJ20220818101205012. 

\bibliographystyle{IEEEtran}
\bibliography{reference}

\begin{thebibliography}{10}
\providecommand{\url}[1]{#1}
\csname url@samestyle\endcsname
\providecommand{\newblock}{\relax}
\providecommand{\bibinfo}[2]{#2}
\providecommand{\BIBentrySTDinterwordspacing}{\spaceskip=0pt\relax}
\providecommand{\BIBentryALTinterwordstretchfactor}{4}
\providecommand{\BIBentryALTinterwordspacing}{\spaceskip=\fontdimen2\font plus
\BIBentryALTinterwordstretchfactor\fontdimen3\font minus \fontdimen4\font\relax}
\providecommand{\BIBforeignlanguage}[2]{{%
\expandafter\ifx\csname l@#1\endcsname\relax
\typeout{** WARNING: IEEEtran.bst: No hyphenation pattern has been}%
\typeout{** loaded for the language `#1'. Using the pattern for}%
\typeout{** the default language instead.}%
\else
\language=\csname l@#1\endcsname
\fi
#2}}
\providecommand{\BIBdecl}{\relax}
\BIBdecl

\bibitem{bammer2003basic}
R.~Bammer, ``Basic principles of diffusion-weighted imaging,'' \emph{European Journal of Radiology}, vol.~45, no.~3, pp. 169--184, 2003.

\bibitem{basser1994estimation}
P.~J. Basser, J.~Mattiello, and D.~LeBihan, ``Estimation of the effective self-diffusion tensor from the nmr spin echo,'' \emph{Journal of Magnetic Resonance, Series B}, vol. 103, no.~3, pp. 247--254, 1994.

\bibitem{le2001diffusion}
D.~Le~Bihan, J.-F. Mangin, C.~Poupon, C.~A. Clark, S.~Pappata, N.~Molko, and H.~Chabriat, ``Diffusion tensor imaging: concepts and applications,'' \emph{Journal of Magnetic Resonance Imaging: An Official Journal of the International Society for Magnetic Resonance in Medicine}, vol.~13, no.~4, pp. 534--546, 2001.

\bibitem{alexander2007diffusion}
A.~L. Alexander, J.~E. Lee, M.~Lazar, and A.~S. Field, ``Diffusion tensor imaging of the brain,'' \emph{Neurotherapeutics}, vol.~4, no.~3, pp. 316--329, 2007.

\bibitem{jiang2006dtistudio}
H.~Jiang, P.~C. Van~Zijl, J.~Kim, G.~D. Pearlson, and S.~Mori, ``Dtistudio: resource program for diffusion tensor computation and fiber bundle tracking,'' \emph{Computer Methods and Programs in Biomedicine}, vol.~81, no.~2, pp. 106--116, 2006.

\bibitem{wilmskoetter2022language}
J.~Wilmskoetter, X.~He, L.~Caciagli, J.~H. Jensen, B.~Marebwa, K.~A. Davis, J.~Fridriksson, A.~Basilakos, L.~P. Johnson, C.~Rorden \emph{et~al.}, ``Language recovery after brain injury: a structural network control theory study,'' \emph{Journal of Neuroscience}, vol.~42, no.~4, pp. 657--669, 2022.

\bibitem{chen2023abnormal}
Y.~Chen, Y.~Wang, Z.~Song, Y.~Fan, T.~Gao, and X.~Tang, ``Abnormal white matter changes in alzheimer's disease based on diffusion tensor imaging: A systematic review,'' \emph{Ageing Research Reviews}, p. 101911, 2023.

\bibitem{harrison2020imaging}
J.~R. Harrison, S.~Bhatia, Z.~X. Tan, A.~Mirza-Davies, H.~Benkert, C.~M. Tax, and D.~K. Jones, ``Imaging alzheimer's genetic risk using diffusion mri: A systematic review,'' \emph{NeuroImage: Clinical}, vol.~27, p. 102359, 2020.

\bibitem{ferreira2014vivo}
P.~F. Ferreira, P.~J. Kilner, L.-A. McGill, S.~Nielles-Vallespin, A.~D. Scott, S.~Y. Ho, K.~P. McCarthy, M.~M. Haba, T.~F. Ismail, P.~D. Gatehouse \emph{et~al.}, ``In vivo cardiovascular magnetic resonance diffusion tensor imaging shows evidence of abnormal myocardial laminar orientations and mobility in hypertrophic cardiomyopathy,'' \emph{Journal of Cardiovascular Magnetic Resonance}, vol.~16, no.~1, p.~87, 2014.

\bibitem{jones2004effect}
D.~K. Jones, ``The effect of gradient sampling schemes on measures derived from diffusion tensor mri: a monte carlo study,'' \emph{Magnetic Resonance in Medicine: An Official Journal of the International Society for Magnetic Resonance in Medicine}, vol.~51, no.~4, pp. 807--815, 2004.

\bibitem{jones2013white}
D.~K. Jones, T.~R. Kn{\"o}sche, and R.~Turner, ``White matter integrity, fiber count, and other fallacies: the do's and don'ts of diffusion mri,'' \emph{Neuroimage}, vol.~73, pp. 239--254, 2013.

\bibitem{landman2007effects}
B.~A. Landman, J.~A. Farrell, C.~K. Jones, S.~A. Smith, J.~L. Prince, and S.~Mori, ``Effects of diffusion weighting schemes on the reproducibility of dti-derived fractional anisotropy, mean diffusivity, and principal eigenvector measurements at 1.5 t,'' \emph{Neuroimage}, vol.~36, no.~4, pp. 1123--1138, 2007.

\bibitem{fick2016mapl}
R.~H. Fick, D.~Wassermann, E.~Caruyer, and R.~Deriche, ``Mapl: Tissue microstructure estimation using laplacian-regularized map-mri and its application to hcp data,'' \emph{NeuroImage}, vol. 134, pp. 365--385, 2016.

\bibitem{jones2014gaussian}
D.~K. Jones, ``Gaussian modeling of the diffusion signal,'' in \emph{Diffusion MRI}.\hskip 1em plus 0.5em minus 0.4em\relax Elsevier, 2014, pp. 87--104.

\bibitem{basu2006rician}
S.~Basu, T.~Fletcher, and R.~Whitaker, ``Rician noise removal in diffusion tensor mri,'' in \emph{Medical Image Computing and Computer-Assisted Intervention--MICCAI 2006: 9th International Conference, Copenhagen, Denmark, October 1-6, 2006. Proceedings, Part I 9}.\hskip 1em plus 0.5em minus 0.4em\relax Springer, 2006, pp. 117--125.

\bibitem{aksoy2011real}
M.~Aksoy, C.~Forman, M.~Straka, S.~Skare, S.~Holdsworth, J.~Hornegger, and R.~Bammer, ``Real-time optical motion correction for diffusion tensor imaging,'' \emph{Magnetic Resonance in Medicine}, vol.~66, no.~2, pp. 366--378, 2011.

\bibitem{jeon2018peripheral}
T.~Jeon, M.~M. Fung, K.~M. Koch, E.~T. Tan, and D.~B. Sneag, ``Peripheral nerve diffusion tensor imaging: overview, pitfalls, and future directions,'' \emph{Journal of Magnetic Resonance Imaging}, vol.~47, no.~5, pp. 1171--1189, 2018.

\bibitem{menzel2011accelerated}
M.~I. Menzel, E.~T. Tan, K.~Khare, J.~I. Sperl, K.~F. King, X.~Tao, C.~J. Hardy, and L.~Marinelli, ``Accelerated diffusion spectrum imaging in the human brain using compressed sensing,'' \emph{Magnetic Resonance in Medicine}, vol.~66, no.~5, pp. 1226--1233, 2011.

\bibitem{zhu2017direct}
Y.~Zhu, X.~Peng, Y.~Wu, E.~X. Wu, L.~Ying, X.~Liu, H.~Zheng, and D.~Liang, ``Direct diffusion tensor estimation using a model-based method with spatial and parametric constraints,'' \emph{Medical Physics}, vol.~44, no.~2, pp. 570--580, 2017.

\bibitem{huang2019accelerating}
J.~Huang, L.~Wang, C.~Chu, W.~Liu, and Y.~Zhu, ``Accelerating cardiac diffusion tensor imaging combining local low-rank and 3d tv constraint,'' \emph{Magnetic Resonance Materials in Physics, Biology and Medicine}, vol.~32, pp. 407--422, 2019.

\bibitem{teh2020improved}
I.~Teh, D.~McClymont, E.~Carruth, J.~Omens, A.~McCulloch, and J.~E. Schneider, ``Improved compressed sensing and super-resolution of cardiac diffusion mri with structure-guided total variation,'' \emph{Magnetic Resonance in Medicine}, vol.~84, no.~4, pp. 1868--1880, 2020.

\bibitem{varela2023single}
G.~Varela-Mattatall, P.~I. Dubovan, T.~Santini, K.~M. Gilbert, R.~S. Menon, and C.~A. Baron, ``Single-shot spiral diffusion-weighted imaging at 7t using expanded encoding with compressed sensing,'' \emph{Magnetic Resonance in Medicine}, vol.~90, no.~2, pp. 615--623, 2023.

\bibitem{waters2011sparcs}
A.~Waters, A.~Sankaranarayanan, and R.~Baraniuk, ``Sparcs: Recovering low-rank and sparse matrices from compressive measurements,'' \emph{Advances in Neural Information Processing Systems}, vol.~24, 2011.

\bibitem{jensen2005diffusional}
J.~H. Jensen, J.~A. Helpern, A.~Ramani, H.~Lu, and K.~Kaczynski, ``Diffusional kurtosis imaging: the quantification of non-gaussian water diffusion by means of magnetic resonance imaging,'' \emph{Magnetic Resonance in Medicine: An Official Journal of the International Society for Magnetic Resonance in Medicine}, vol.~53, no.~6, pp. 1432--1440, 2005.

\bibitem{golkov2016q}
V.~Golkov, A.~Dosovitskiy, J.~I. Sperl, M.~I. Menzel, M.~Czisch, P.~S{\"a}mann, T.~Brox, and D.~Cremers, ``Q-space deep learning: twelve-fold shorter and model-free diffusion mri scans,'' \emph{IEEE Transactions on Medical Imaging}, vol.~35, no.~5, pp. 1344--1351, 2016.

\bibitem{ronneberger2015u}
O.~Ronneberger, P.~Fischer, and T.~Brox, ``U-net: Convolutional networks for biomedical image segmentation,'' in \emph{Medical Image Computing and Computer-assisted Intervention--MICCAI 2015: 18th International Conference, Munich, Germany, October 5-9, 2015, proceedings, part III 18}.\hskip 1em plus 0.5em minus 0.4em\relax Springer, 2015, pp. 234--241.

\bibitem{tian2020deepdti}
Q.~Tian, B.~Bilgic, Q.~Fan, C.~Liao, C.~Ngamsombat, Y.~Hu, T.~Witzel, K.~Setsompop, J.~R. Polimeni, and S.~Y. Huang, ``Deepdti: High-fidelity six-direction diffusion tensor imaging using deep learning,'' \emph{NeuroImage}, vol. 219, p. 117017, 2020.

\bibitem{li2021superdti}
H.~Li, Z.~Liang, C.~Zhang, R.~Liu, J.~Li, W.~Zhang, D.~Liang, B.~Shen, X.~Zhang, Y.~Ge \emph{et~al.}, ``Superdti: Ultrafast dti and fiber tractography with deep learning,'' \emph{Magnetic Resonance in Medicine}, vol.~86, no.~6, pp. 3334--3347, 2021.

\bibitem{karimi2022diffusion}
D.~Karimi and A.~Gholipour, ``Diffusion tensor estimation with transformer neural networks,'' \emph{Artificial Intelligence in Medicine}, vol. 130, p. 102330, 2022.

\bibitem{wang2024ultrafast}
J.~Wang, Z.~Chen, C.~Cai, and S.~Cai, ``Ultrafast diffusion tensor imaging based on deep learning and multi-slice information sharing,'' \emph{Physics in Medicine \& Biology}, vol.~69, no.~3, p. 035011, 2024.

\bibitem{li2023diffusion}
Z.~Li, Q.~Fan, B.~Bilgic, G.~Wang, W.~Wu, J.~R. Polimeni, K.~L. Miller, S.~Y. Huang, and Q.~Tian, ``Diffusion mri data analysis assisted by deep learning synthesized anatomical images (deepanat),'' \emph{Medical Image Analysis}, vol.~86, p. 102744, 2023.

\bibitem{liu2023accelerated}
S.~Liu, Y.~Liu, X.~Xu, R.~Chen, D.~Liang, Q.~Jin, H.~Liu, G.~Chen, and Y.~Zhu, ``Accelerated cardiac diffusion tensor imaging using deep neural network,'' \emph{Physics in Medicine \& Biology}, vol.~68, no.~2, p. 025008, 2023.

\bibitem{huff2021interpretation}
D.~T. Huff, A.~J. Weisman, and R.~Jeraj, ``Interpretation and visualization techniques for deep learning models in medical imaging,'' \emph{Physics in Medicine \& Biology}, vol.~66, no.~4, p. 04TR01, 2021.

\bibitem{dhariwal2021diffusion}
P.~Dhariwal and A.~Nichol, ``Diffusion models beat gans on image synthesis,'' \emph{Advances in Neural Information Processing Systems}, vol.~34, pp. 8780--8794, 2021.

\bibitem{cao2024survey}
H.~Cao, C.~Tan, Z.~Gao, Y.~Xu, G.~Chen, P.-A. Heng, and S.~Z. Li, ``A survey on generative diffusion models,'' \emph{IEEE Transactions on Knowledge and Data Engineering}, 2024.

\bibitem{yang2023diffusion}
L.~Yang, Z.~Zhang, Y.~Song, S.~Hong, R.~Xu, Y.~Zhao, W.~Zhang, B.~Cui, and M.-H. Yang, ``Diffusion models: A comprehensive survey of methods and applications,'' \emph{ACM Computing Surveys}, vol.~56, no.~4, pp. 1--39, 2023.

\bibitem{ho2020denoising}
J.~Ho, A.~Jain, and P.~Abbeel, ``Denoising diffusion probabilistic models,'' \emph{Advances in Neural Information Processing Systems}, vol.~33, pp. 6840--6851, 2020.

\bibitem{song2020score}
\BIBentryALTinterwordspacing
Y.~Song, J.~Sohl{-}Dickstein, D.~P. Kingma, A.~Kumar, S.~Ermon, and B.~Poole, ``Score-based generative modeling through stochastic differential equations,'' in \emph{9th International Conference on Learning Representations, {ICLR} 2021, Virtual Event, Austria, May 3-7, 2021}.\hskip 1em plus 0.5em minus 0.4em\relax OpenReview.net, 2021. [Online]. Available: \url{https://openreview.net/forum?id=PxTIG12RRHS}
\BIBentrySTDinterwordspacing

\bibitem{chung2022score}
H.~Chung and J.~C. Ye, ``Score-based diffusion models for accelerated mri,'' \emph{Medical Image Analysis}, vol.~80, p. 102479, 2022.

\bibitem{song2021solving}
\BIBentryALTinterwordspacing
Y.~Song, L.~Shen, L.~Xing, and S.~Ermon, ``Solving inverse problems in medical imaging with score-based generative models,'' in \emph{The Tenth International Conference on Learning Representations, {ICLR} 2022, Virtual Event, April 25-29, 2022}.\hskip 1em plus 0.5em minus 0.4em\relax OpenReview.net, 2022. [Online]. Available: \url{https://openreview.net/forum?id=vaRCHVj0uGI}
\BIBentrySTDinterwordspacing

\bibitem{wolleb2022diffusion}
J.~Wolleb, F.~Bieder, R.~Sandk{\"u}hler, and P.~C. Cattin, ``Diffusion models for medical anomaly detection,'' in \emph{International Conference on Medical image computing and computer-assisted intervention}.\hskip 1em plus 0.5em minus 0.4em\relax Springer, 2022, pp. 35--45.

\bibitem{yoon2023sadm}
J.~S. Yoon, C.~Zhang, H.-I. Suk, J.~Guo, and X.~Li, ``Sadm: Sequence-aware diffusion model for longitudinal medical image generation,'' in \emph{International Conference on Information Processing in Medical Imaging}.\hskip 1em plus 0.5em minus 0.4em\relax Springer, 2023, pp. 388--400.

\bibitem{wang2024two}
W.~Wang, Z.-X. Cui, G.~Cheng, C.~Cao, X.~Xu, Z.~Liu, H.~Wang, Y.~Qi, D.~Liang, and Y.~Zhu, ``A two-stage generative model with cyclegan and joint diffusion for mri-based brain tumor detection,'' \emph{IEEE Journal of Biomedical and Health Informatics}, 2024.

\bibitem{cao2024high}
C.~Cao, Z.-X. Cui, Y.~Wang, S.~Liu, T.~Chen, H.~Zheng, D.~Liang, and Y.~Zhu, ``High-frequency space diffusion model for accelerated mri,'' \emph{IEEE Transactions on Medical Imaging}, 2024.

\bibitem{song2020improved}
\BIBentryALTinterwordspacing
Y.~Song and S.~Ermon, ``Improved techniques for training score-based generative models,'' in \emph{Advances in Neural Information Processing Systems 33: Annual Conference on Neural Information Processing Systems 2020, NeurIPS 2020, December 6-12, 2020, virtual}, H.~Larochelle, M.~Ranzato, R.~Hadsell, M.~Balcan, and H.~Lin, Eds., 2020. [Online]. Available: \url{https://proceedings.neurips.cc/paper/2020/hash/92c3b916311a5517d9290576e3ea37ad-Abstract.html}
\BIBentrySTDinterwordspacing

\bibitem{elam2021human}
J.~S. Elam, M.~F. Glasser, M.~P. Harms, S.~N. Sotiropoulos, J.~L. Andersson, G.~C. Burgess, S.~W. Curtiss, R.~Oostenveld, L.~J. Larson-Prior, J.-M. Schoffelen \emph{et~al.}, ``The human connectome project: a retrospective,'' \emph{NeuroImage}, vol. 244, p. 118543, 2021.

\bibitem{maggioni2012nonlocal}
M.~Maggioni, V.~Katkovnik, K.~Egiazarian, and A.~Foi, ``Nonlocal transform-domain filter for volumetric data denoising and reconstruction,'' \emph{IEEE Transactions on Image Processing}, vol.~22, no.~1, pp. 119--133, 2012.

\bibitem{zhu2017unpaired}
J.-Y. Zhu, T.~Park, P.~Isola, and A.~A. Efros, ``Unpaired image-to-image translation using cycle-consistent adversarial networks,'' in \emph{Proceedings of The IEEE International Conference on Computer Vision}, 2017, pp. 2223--2232.

\bibitem{loizillon2024leveraging}
S.~Loizillon, S.~Mabille, S.~Bottani, Y.~Jacob, A.~Maire, S.~Str{\"o}er, D.~Dormont, O.~Colliot, and N.~Burgos, ``Leveraging noise and contrast simulation for the automatic quality control of routine clinical t1-weighted brain mri,'' in \emph{Medical Imaging 2024: Image Processing}, vol. 12926.\hskip 1em plus 0.5em minus 0.4em\relax SPIE, 2024, pp. 322--326.

\bibitem{loizillon2024detecting}
S.~Loizillon, Y.~Jacob, M.~Aur{\'e}lien, D.~Dormont, O.~Colliot, N.~Burgos, A.~S. Group \emph{et~al.}, ``Detecting brain anomalies in clinical routine with the \(\beta\)-vae: Feasibility study on age-related white matter hyperintensities,'' in \emph{Medical Imaging with Deep Learning}, 2024.

\bibitem{robertson2018effects}
J.~Robertson, J.~Urban, J.~Stitzel, and B.~E. Treeby, ``The effects of image homogenisation on simulated transcranial ultrasound propagation,'' \emph{Physics in Medicine \& Biology}, vol.~63, no.~14, p. 145014, 2018.

\bibitem{youm2016image}
G.-Y. Youm, S.-H. Bae, and M.~Kim, ``Image super-resolution based on convolution neural networks using multi-channel input,'' in \emph{2016 IEEE 12th Image, Video, and Multidimensional Signal Processing Workshop (IVMSP)}.\hskip 1em plus 0.5em minus 0.4em\relax IEEE, 2016, pp. 1--5.

\bibitem{souza2020dual}
R.~Souza, M.~Bento, N.~Nogovitsyn, K.~J. Chung, W.~Loos, R.~M. Lebel, and R.~Frayne, ``Dual-domain cascade of u-nets for multi-channel magnetic resonance image reconstruction,'' \emph{Magnetic Resonance Imaging}, vol.~71, pp. 140--153, 2020.

\bibitem{lyu2022conversion}
Q.~Lyu and G.~Wang, ``Conversion between ct and mri images using diffusion and score-matching models,'' 2022.

\bibitem{li2024dimond}
Z.~Li, Z.~Li, B.~Bilgic, H.-H. Lee, K.~Ying, S.~Y. Huang, H.~Liao, and Q.~Tian, ``Dimond: Diffusion model optimization with deep learning,'' \emph{Advanced Science}, p. 2307965, 2024.

\bibitem{chen2022image}
R.~Chen, D.~Pu, Y.~Tong, and M.~Wu, ``Image-denoising algorithm based on improved k-singular value decomposition and atom optimization,'' \emph{CAAI Transactions on Intelligence Technology}, vol.~7, no.~1, pp. 117--127, 2022.

\bibitem{luo2023refusion}
Z.~Luo, F.~K. Gustafsson, Z.~Zhao, J.~Sj{\"o}lund, and T.~B. Sch{\"o}n, ``Refusion: Enabling large-size realistic image restoration with latent-space diffusion models,'' in \emph{Proceedings of The IEEE/CVF Conference on Computer Vision and Pattern Recognition}, 2023, pp. 1680--1691.

\bibitem{kim2024adaptive}
J.~Kim and H.~Park, ``Adaptive latent diffusion model for 3d medical image to image translation: Multi-modal magnetic resonance imaging study,'' in \emph{Proceedings of the IEEE/CVF Winter Conference on Applications of Computer Vision}, 2024, pp. 7604--7613.

\bibitem{zhang2024dosediff}
Y.~Zhang, C.~Li, L.~Zhong, Z.~Chen, W.~Yang, and X.~Wang, ``Dosediff: Distance-aware diffusion model for dose prediction in radiotherapy,'' \emph{IEEE Transactions on Medical Imaging}, 2024.

\end{thebibliography}

\end{document}